\documentclass[twocolumn,english,aps,prl,floatfix,tightenlines]{revtex4-1}
\usepackage[T1]{fontenc}
\usepackage[latin9]{inputenc}
\usepackage{amsmath}
\usepackage{amssymb}
\usepackage{graphicx}
\usepackage{esint}

\makeatletter

\@ifundefined{textcolor}{}
{%
 \definecolor{BLACK}{gray}{0}
 \definecolor{WHITE}{gray}{1}
 \definecolor{RED}{rgb}{1,0,0}
 \definecolor{GREEN}{rgb}{0,1,0}
 \definecolor{BLUE}{rgb}{0,0,1}
 \definecolor{CYAN}{cmyk}{1,0,0,0}
 \definecolor{MAGENTA}{cmyk}{0,1,0,0}
 \definecolor{YELLOW}{cmyk}{0,0,1,0}
}

\usepackage{babel}
\usepackage{times}
\usepackage{hyperref}

\newcommand{\tr}{\mathrm{tr}}

\newcommand{\1}{\leavevmode{\rm 1\ifmmode\mkern  -4.8mu\else\kern -.3em\fi I}}
\usepackage{url}

\makeatother

\usepackage{babel}
\begin{document}

\title{The recurrence time in quantum mechanics}

\author{Lorenzo Campos Venuti}

\affiliation{Department of Physics and Astronomy and Center for Quantum Information
Science \& Technology, University of Southern California, Los Angeles,
CA 90089-0484, USA}
\begin{abstract}
Generic quantum systems --as much as their classical counterparts--
pass arbitrarily close to their initial state after sufficiently long
time. Here we provide an essentially exact computation of such recurrence
times for generic non-integrable quantum models. The result is a universal
function which depends on just two parameters, an energy scale and
the effective dimension of the system. As a by-product we prove that
the density of orthogonalization times is zero if at least nine levels
are populated and connections with the quantum speed limit are discussed.
We also extend our results to integrable, quasi-free fermions. For
generic systems the recurrence time is generally doubly exponential
in the system volume whereas for the integrable case the dependence
is only exponential. The recurrence time can be decreased by several
orders of magnitude by performing a small quench  close to a quantum
critical point. This setup may lead to the experimental observation
of such \emph{fast} recurrences. 
\end{abstract}
\maketitle

\section{Introduction}

The idea that the world will repeat itself after some time is an ancient
one and appears in many philosophies and cultures. Babylonians called
the Great Year the time needed for the planets to return to their
initial conditions \cite{cagni_poem_1977,marinus_van_der_sluijs_possible_2005}
\footnote{Since the planets' orbital period are not commensurate their estimate
for the Great Year kept increasing.%
}. However it was only in the end of the XIX century that Henri Poincar\'e
rigorously proved the existence of a recurrence time for a certain
description of reality \cite{poincare_sur_1890}. The beauty and the
power of Poincar\'e's result lie in its simplicity and in its wide
range of applicability. The only hypotheses of Poincar\'e recurrence
theorem are that 1) the (phase) space available to the dynamics is
bounded in volume; and 2) the dynamical flow $g^{t}$ preserves the
volumes %
\footnote{Note that there is a hidden hypothesis that the system is described
by a classical dynamical equation such that the jargon in 1) and 2)
makes sense.%
}. Under such hypotheses the system returns arbitrarily close to its
initial conditions after sufficiently long time. Hypothesis 2) is
satisfied if we believe a classical, Hamiltonian, description of the
world thanks to Liouville's theorem. Hypothesis 1) holds true if,
for instance, the motion is restricted to a bounded region of space
because of conservation of energy \cite{gibbs_elementary_1902}. Poincar\'e
recurrence theorem has philosophical and counterintuitive implications.
 For example --if we believe a conservative, classical dynamics--
the theorem predicts that, if we let two different gases mix by removing
a barrier, there will exist a time where we will see the two gases
separate again. Indeed Zermelo used Poincar\'e result's to argue
that Boltzmann's formula for the entropy should decrease after some
sufficiently long time, thus violating the claim that entropy always
increases \cite{zermelo_ueber_1896}. The reason why we never see
the two gases nicely separate back again is due to the astronomically
large times needed to observe such recurrences. Poincar\'e's argument
does not give any indication on how to estimate such recurrence times.
In order to do that one must generally have stronger assumptions.
The very first estimate of recurrence time has been given by Boltzmann
himself in reply to Zermelo's criticism \cite{boltzmann_entgegnung_1896}.
Boltzmann estimated that the time needed for a $\mathrm{cm}^{3}$
of gas to go back to its initial state, has of the order of many trillion
of digits. 

In more recent years the estimation of such Poincar\'e recurrences
became an active research topic belonging to the field of dynamical
systems \cite{barreira_poincare_2006}. The first estimation of return
times appeared in \cite{hemmer_recurrence_1958} for a system of harmonic
oscillators. More precisely in Ref.~\cite{hemmer_recurrence_1958}
the average recurrence has been computed. It is defined by $T_{R}=\lim_{r\to\infty}t_{r}/r,$
where $t_{r}$ is the location of the $r$-th recurrence. As noted
by Kac in \cite{kac_remark_1959}, the result of Ref.~\cite{hemmer_recurrence_1958}
can be obtained almost immediately thanks to the following formula
of Smoluchowski, valid for a discrete evolution,
\begin{equation}
T_{R}=\tau\frac{\mu(S)-\mu(A)}{\mu(A)-\mu(g^{\tau}A\cap A)}.\label{eq:smoluchowski}
\end{equation}
The above formula gives the average time taken to return in set $A$
given that we started from a point of the set. In Eq.~(\ref{eq:smoluchowski})
$\mu(S)$ is the (finite) invariant measure of the total available
phase space and $g^{\tau}$ is the mapping advancing one unit of time
$\tau$ (see e.g.~\cite{arnold_ordinary_1978}). The result for continuous
evolution is obtained taking the limit $\tau\to0$. Eq.~(\ref{eq:smoluchowski})
can readily be applied to a classical integrable system. The assumption
of finiteness of the available phase space (more precisely the manifold
is assumed to be compact and simply connected), implies that the motion
takes place on a $N$ dimensional torus $\mathbb{T}^{N}=\{(\phi_{1},\ldots,\phi_{N})\mod2\pi\}$,
where $N$ is the number of degrees of freedom \cite{arnold_mathematical_1989}.
The map $g^{t}$ is given by $g^{t}\boldsymbol{\phi}=(\phi_{1}+\omega_{1}t,\ldots,\phi_{N}+\omega_{N}t)$.
If the frequencies $\{\omega_{i}\}$ are rationally independent (RI,
i.e.~linearly independent over the field of the rationals), time
averages are equivalent to phase space averages on the torus and Eq.~(\ref{eq:smoluchowski})
applies. In this case, if $A$ is taken to be the angular interval
$(\Delta\phi_{1},\ldots,\Delta\phi_{N})$, Eq.~(\ref{eq:smoluchowski})
gives, after taking the $\tau\to0$ limit, 
\begin{equation}
T_{R}=\frac{\left(\prod_{j=1}^{N}\frac{2\pi}{\Delta\phi_{j}}-1\right)}{\sum_{j=1}^{N}\frac{\omega_{j}}{\Delta\phi_{j}}}.\label{eq:TR_torus}
\end{equation}
For example, taking for simplicity $\Delta\phi_{j}=\epsilon$ we obtain
$T_{R}\simeq\epsilon e^{\ln(2\pi/\epsilon)N}/\sum_{j=1}^{N}\omega_{j}$. 

We know however reality is ultimately quantum. Does a recurrence theorem
applies also to the quantum world? That this is the case was shown
in Ref.~\cite{bocchieri_quantum_1957} but no actual estimate of
recurrence time was given. Some approximate estimates for some specific
initial states where given in \cite{peres_recurrence_1982,bhattacharyya_estimates_1986}.
In this paper we solve this problem in an essentially exact way. 

\begin{figure}
\begin{centering}
\vspace{3mm}
\par\end{centering}

\noindent \begin{centering}
\includegraphics[width=4cm,height=3cm]{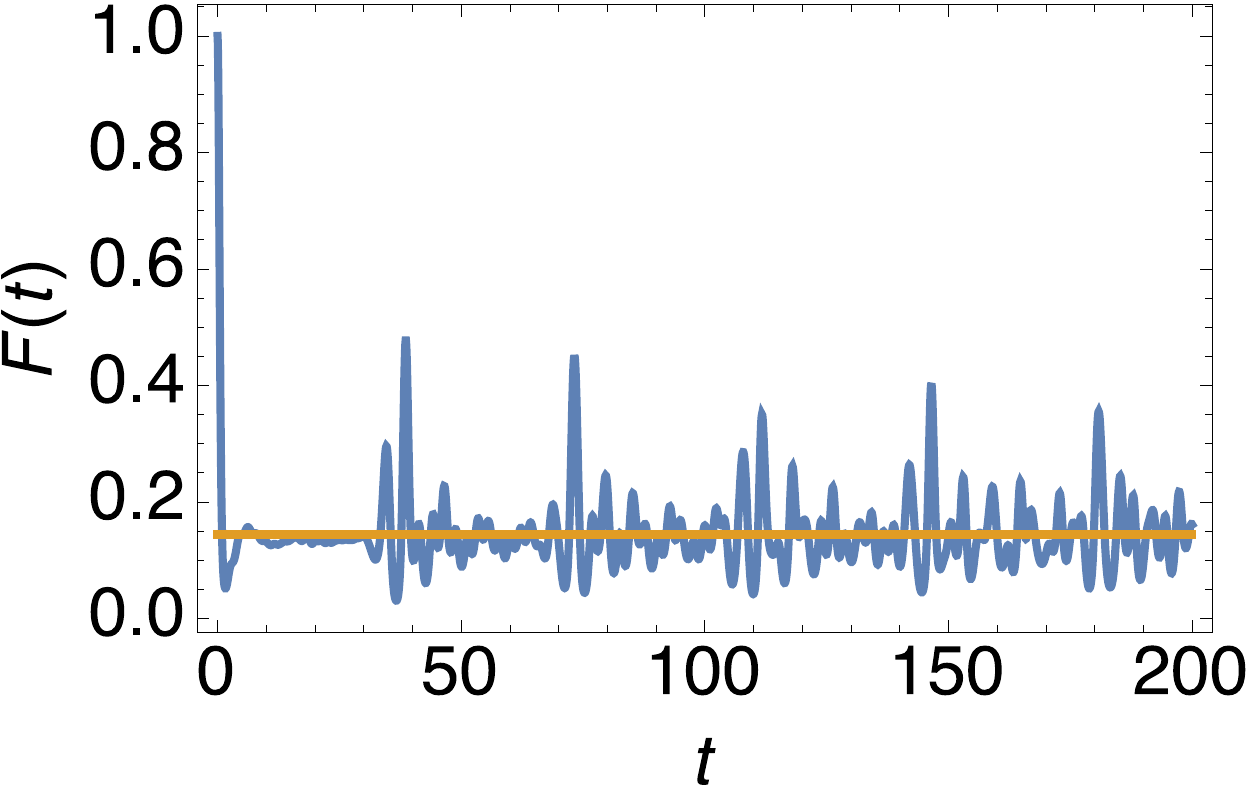}\includegraphics[width=4.2cm,height=3cm]{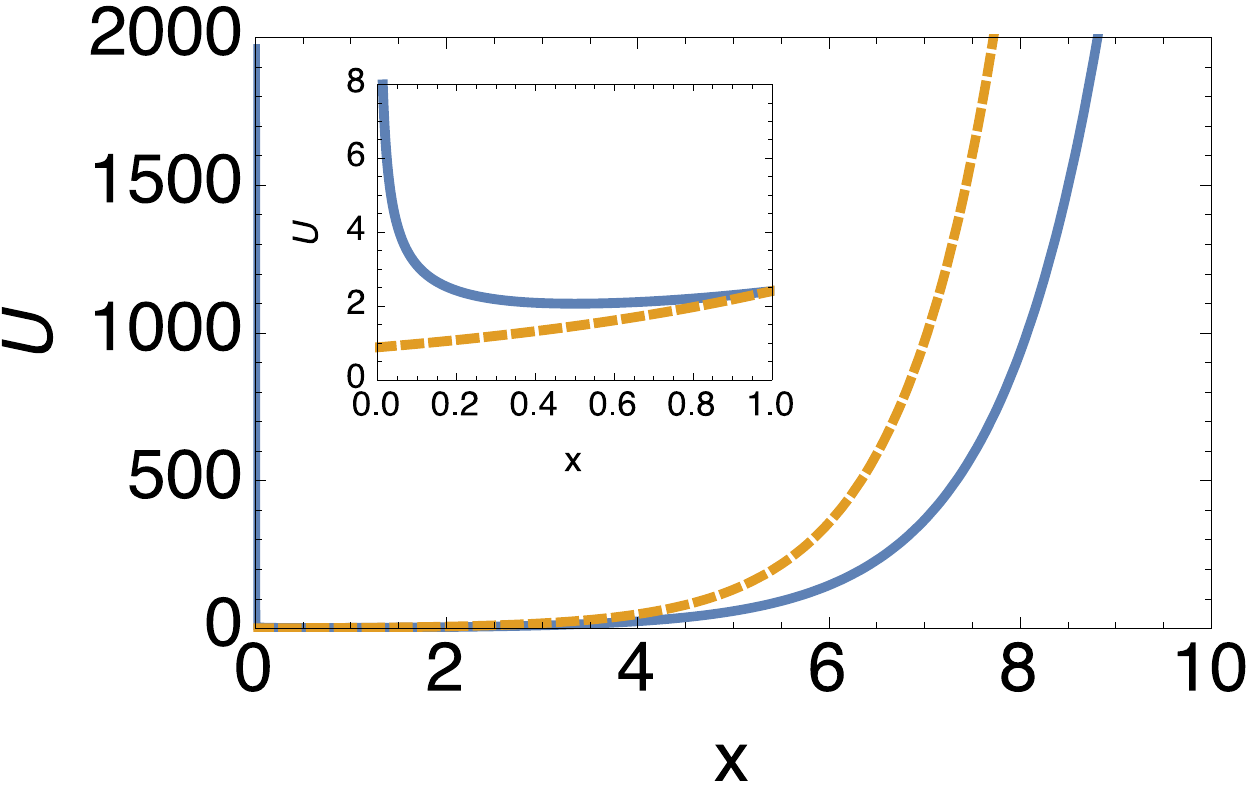}
\par\end{centering}

\protect\caption{Left panel: typical behavior of the fidelity for a many body system.
The system is given by Hamiltonian Eq.~(\ref{eq:TAM}) with $\kappa=0$
with $L=50$. The horizontal line is the location of the average $\overline{\mathcal{F}}$.
The peaks observed at times approximately integer multiple of the
system size are \emph{revivals} discussed in \cite{campos_venuti_unitary_2010}.
Right panel: universal function $\mathcal{U}(x)=(\sqrt{\pi}/2)e^{x}/\sqrt{x}$.
$\mathcal{U}(x)$ has essentially three regimes: for $x\ll1$, $\mathcal{U}(x)\sim1/\sqrt{x}$,
for $x=O(1)$, $\mathcal{U}(x)\sim O(1)$ whereas for $x\gg1$, $\mathcal{U}(x)\sim e^{x}$
(dashed line). \label{fig:LEt}}
\end{figure}

Let's assume then, from here on, that the world evolves according
to Schr\"{o}dinger equation and that, for simplicity, the initial
state is pure. For the time being we assume that the system's Hamiltonian
$H$ has purely discrete spectrum. This is a crucial assumption on
which we will come back (and weaken it) later. Then $H=\sum_{n}E_{n}\Pi_{n}$
with orthogonal projectors $\Pi_{n}$ and let $|\psi\rangle=|\psi(0)\rangle$
the state of the system at $t=0$. We define $|n\rangle=\Pi_{n}|\psi\rangle/\left\Vert \Pi_{n}|\psi\rangle\right\Vert $.
The state at time $t$ is then $|\psi(t)\rangle=e^{-itH}|\psi\rangle=\sum_{n}\left\Vert \Pi_{n}|\psi\rangle\right\Vert e^{-itE_{n}}|n\rangle$
\footnote{We use throughout units for which $\hbar=1$. %
}. As time goes by $|\psi(t)\rangle$ spans a $d$ dimensional torus
($d$, possibly infinite, is the number of non-zero $\left\Vert \Pi_{n}|\psi\rangle\right\Vert $)
with radii $\left\Vert \Pi_{n}|\psi\rangle\right\Vert $ and angular
frequencies $E_{n}$. If the energies $E_{n}$ are rationally independent
(which is natural to assume) the torus is filled uniformly as time
goes by. It is tempting then to try to recycle the classical result
and use Eq.~(\ref{eq:TR_torus}) to estimate the recurrence time
in the quantum case. The problem with that is that the distance of
points on the torus does not imply any physical distance between states.
For example states $|n\rangle$ with large amplitudes $\left\Vert \Pi_{n}|\psi\rangle\right\Vert $
should count more than states with small amplitudes. A natural distance
between $|\psi(t)\rangle$ and $|\psi\rangle$, instead, is encoded
in the fidelity 
\begin{equation}
\mathcal{F}(t)=\left|\langle\psi|\psi(t)\rangle\right|^{2}.\label{eq:LE}
\end{equation}
The quantity in Eq.~(\ref{eq:LE}) is also known under different
names such as survival probabilities or Loschmidt echo, and it has
been studied in various contexts such as quantum chaos, the theory
of fermi edge singularities, equilibration and decoherence \cite{gorin_dynamics_2006,schotte_tomonagas_1969,silva_statistics_2008,campos_venuti_theory_2015-1,quan_decay_2006}.
The statistical properties of $\mathcal{F}(t)$ when $t$ is seen
as a uniform random variable on the infinite half-line have been investigated
in a series of works \cite{campos_venuti_unitary_2010,campos_venuti_universality_2010,campos_venuti_gaussian_2013,campos_venuti_universal_2014,campos_venuti_theory_2015,campos_venuti_theory_2015-1}.
If the spectrum of $H$ is discrete, $\mathcal{F}(t)$ is an almost
periodic function \cite{besicovitch_almost_1954} and $\mathcal{F}(t)$
evolves indefinitely without ever converging to a limit as $t\to\infty$.
Conversely if the spectrum is purely continuous (or if only this part
of the spectrum is populated) $\lim_{t\to\infty}\mathcal{F}(t)=0$
essentially because of the Riemann--Lebesgue lemma. In this situation
the system does not return close to its initial state and actually
becomes orthogonal to it for long times %
\footnote{For example, for a Gaussian wave-packet of spread $\sigma$ freely
evolving in one dimension, $\mathcal{F}(t)=1/\sqrt{1+(t/\tau)^{2}}$
(where $\tau=4m\sigma/\hbar$ and $m$ is the particle's mass).%
}. Hence only if there is a discrete spectrum, and if the corresponding
probabilities $p_{n}=\left\Vert \Pi_{n}|\psi\rangle\right\Vert ^{2}$
are non-zero, the system has a chance to return close to its initial
state. We henceforth set ourself in this situation. In this case,
typically, $\mathcal{F}(t)$ start from its maximum one at $t=0,$
drops to an average value $\overline{\mathcal{F}}$ in a time that
may well be called equilibration time \cite{campos_venuti_unitary_2010,campos_venuti_equilibration_2013},
and starts oscillating around this value (see Fig.~\ref{fig:LEt}
for a typical plot). We are interested in estimating the time $\mathcal{F}(t)$
will go back to a value $u\in[0,1]$ possibly (but not necessarily)
close to one. Let us proceed as follows. We call $t_{n}$ the $n$-th
solution of the equation $\mathcal{F}(t)=u$ on the half-line $t\ge0$
starting from the left. We define the (average) recurrence time as
$T_{R}(u)=\lim_{n\to\infty}t_{n}/n$.  This number clearly does not
need to coincide with, for instance the first recurrence, but it's
essentially the only quantity that makes physical sense in that --as
we will show-- depends only on a finite number of physically relevant
constants. Estimating the exact recurrences $t_{n}$, besides being
mathematically almost intractable, is physically meaningless. In fact
the exact recurrences $t_{n}$ depend crucially on the energy levels
$E_{n}$ which must be known with \emph{infinite accuracy }\cite{lenstra_factoring_1982,kovacs_efficient_2013,jonckheere_information_2014}\emph{.
}Since the latter\emph{ }seems a rather unphysical presumption we
focus on the computation of $T_{R}(u)$. 

Our strategy will be the following. First we estimate the number of
zeros $N_{T}(u)$ of the equation $\mathcal{F}(t)=u$ in a large segment
$[0,T]$ with $T\gg1$. We will show that, for large $T$, $N_{T}(u)\simeq T\, D(u)$
with a finite density $D(u)$. Then the recurrence time is given by
$T_{R}(u)=\lim_{T\to\infty}T/N_{T}(u)=1/D(u)$. 

We now show how to compute $D(u)$. Similar techniques have been pioneered
by Kac and Rice to study the average number of roots of a random polynomial
or time signal \cite{kac_average_1943,rice_mathematical_1944}.

\section{Non-integrable case}

We assume here the the Hamiltonian $H$ has pure discrete spectrum
and will consider generalizations later. This condition plays an analogous
role as the requirement of a compact phase space for the classical
case. The Hamiltonian has then the following spectral decomposition
$H=\sum_{n}E_{n}\Pi_{n}$ with eigen-projectors $\Pi_{n}$. 

The number of zeros of the function $\mathcal{F}(t)-u$ in the interval
$[0,T]$, is given by the following formula
\begin{equation}
N_{T}(u)=\int_{0}^{T}\delta(\mathcal{F}(t)-u)\left|\mathcal{F}'(t)\right|dt,\label{eq:Nu}
\end{equation}
as can be checked by a change of variable (and with $\mathcal{F}'(t)=d\mathcal{F}(t)/dt$).
For large $T$ we can use $N_{T}(u)=TD(u)+O(T^{0})$ where the density
is given by $D(u)=\lim_{T\to\infty}T^{-1}\int_{0}^{T}\delta(\mathcal{F}(t)-u)\left|\mathcal{F}'(t)\right|dt$
provided $D(u)$ is finite. The density can then be computed with
\begin{equation}
D(u)=\int df'\left|f'\right|P_{\mathcal{F},\mathcal{F}'}(u,f')\label{eq:Du}
\end{equation}
where $P_{\mathcal{F},\mathcal{F}'}(f,f'):=\overline{\delta(\mathcal{F}(t)-f)\delta(\mathcal{F}'(t)-f')}$
is the joint probability distribution of $\mathcal{F}$ and $\mathcal{F}'$,
and overline indicates infinite time average, i.e.~$\overline{a(t)}:=\lim_{T\to\infty}T^{-1}\int_{0}^{T}a(t)dt$.
For the next step we assume that the energies $E_{n}$ are rationally
independent, i.e.~linearly independent over the field of rationals.
This is a very reasonable assumptions for generic non-integrable system.
Thanks to rational independence we can express time averages as phase
space averages over a large multi-dimensional torus \cite{arnold_mathematical_1989}.
The steps are detailed in the Appendix \ref{sec:zero_density}. The
main ingredient is given by the fact a certain distribution related
to $P_{\mathcal{F},\mathcal{F}'}$, becomes Gaussian in the limit
of large dimensionality. 

The final result for the recurrence time is surprisingly simple
\begin{equation}
T_{R}(u)=\frac{\sqrt{\pi}}{2}\frac{1}{\Delta E}\sqrt{\frac{\overline{\mathcal{F}}}{u}}e^{u/\overline{\mathcal{F}}},\label{eq:TR}
\end{equation}
where $\overline{\mathcal{F}}=\sum_{j}p_{j}^{2}$ and $\Delta E^{2}$
is the variance of the energy $\{E_{n}\}$ with respect to the ensemble
$\nu_{n}=p_{n}^{2}/\sum_{n}p_{n}^{2}$, i.e.~$\Delta E^{2}=[\sum_{j}p_{j}^{2}E_{j}^{2}/\overline{\mathcal{F}}-(\sum_{j}p_{j}^{2}E_{j}/\overline{\mathcal{F}})^{2}]$.
The factor $1/\Delta E$ is essentially needed to reproduce the correct
dimensionality whereas the main dependence is through a universal
function of $x=u/\overline{\mathcal{F}}$, $\mathcal{U}(x)=(\sqrt{\pi}/2)e^{x}/\sqrt{x}$
(see Fig.~\ref{fig:LEt} right panel). Since, forgetting the size
dependence, $\Delta E\sim J$ where $J$ is some energy scale of $H$,
a good approximation of Eq.~(\ref{eq:TR}) is given by $T_{R}\sim J^{-1}\sqrt{\overline{\mathcal{F}}/u}e^{u/\overline{\mathcal{F}}}$.
Several considerations can be drawn on the basis of Eq.~(\ref{eq:TR}):

\textbf{1)} For macroscopic systems the recurrence time is, in general,
\emph{doubly exponential }in the system's volume. Indeed, in the typical
situation, the average fidelity is exponentially small in the system
volume, $\overline{\mathcal{F}}\sim e^{-\alpha V}$ with $\alpha$
positive constant (see e.g.~\cite{campos_venuti_gaussian_2013})
\footnote{Here and in the following we mean the dimensionless volume normalized
by the size of the unit cell, i.e.~the total number of cells. %
}. Sufficient conditions are i) the Hamiltonian is local and extensive,
meaning $H$ can be written as $H=\sum_{x}h(x)$, with $h(x)$ locally
supported operators; together with ii) the initial state is exponentially
clustering, i.e.~the state $\langle\bullet\rangle$ satisfies $\langle h(x)h(y)\rangle-\langle h(x)\rangle\langle h(y)\rangle=O\left(e^{-|x-y|/\xi}\right)$.
Hence, under these very general conditions, one has readily $T_{R}(u)\sim J^{-1}e^{ue^{\alpha V}}$
(for $u\gg\overline{\mathcal{F}}$).

\textbf{2)} It is tempting to relate the recurrence time to the concept
of entropy. With the aid of the equilibrium state $\overline{\rho}:=\overline{|\psi(t)\rangle\langle\psi(t)|}$
one can write $H_{2}(\overline{\rho})=-\ln\overline{\mathcal{F}}$,
the 2-R\'{e}nyi entropy of $\overline{\rho}$. Recall the entropy
inequality $H_{2}(\overline{\rho})\le S(\overline{\rho})$ where $S$
is the Von Neumann entropy of $\overline{\rho}$, $S(\overline{\rho})=-\tr\overline{\rho}\ln\overline{\rho}$.
The inequality becomes tight in the limit of totally mixed $\overline{\rho}$,
i.e.~when $\overline{\mathcal{F}}$ is small. Since, as we have shown,
typically $\overline{\mathcal{F}}\sim e^{-\alpha V}$ a possible proxy
to the recurrence time (for $u\gg\overline{\mathcal{F}}$) is given
by (omitting pre-factors) $T_{R}\sim J^{-1}e^{ue^{S}}$, a formula
sometimes found in the literature (see e.g., Ref.~\cite{page_information_1994}).

\textbf{3)} Equation (\ref{eq:TR}) and its inverse $D(u)=1/T_{R}(u)$
should be handled with care at the extrema $u=0$ or $u=1$. For example
Eq.~(\ref{eq:TR}) predicts a \emph{finite} density of exact recurrences
($u=1$) but a zero density of orthogonalization times (i.e.~times
for which $\mathcal{F}(t)=0$). This is quite surprising given the
fact that typically $\mathcal{F}(t)$ spends much of the time close
to its average $\overline{\mathcal{F}}$ which in turn is very close
to zero. One can ask wether these results are correct or are a consequence
of the Gaussian approximation used to derive (\ref{eq:TR}). The tiny
density of exact recurrences (for $u=1$) predicted by Eq.~(\ref{eq:TR}),
is indeed a consequence of the Gaussian approximation. In fact, for
rationally independent energies the only solution of $\mathcal{F}(t)=1$
is at $t=0$  (see Appendix \ref{sec:LEeq1} for a proof). Instead,
using this approach without the Gaussian approximation we have been
able to prove that indeed the density of orthogonalization times is
zero (see Appendix \ref{sec:zero_density}), provided at least nine
energy levels are populated and the energies are RI.  

\textbf{4)} For $u$ close to 1 the roots of $\mathcal{F}(t)=u$ are
very close to a maximum of $\mathcal{F}(t)$, hence they must come
in close pairs. This means that essentially, for $u\approx1$ a more
precise estimate of the recurrence time is given by $2T_{R}(u)$.
More importantly, on the basis of numerical simulations, we observed
that, for $u$ close to one, the solutions of $\mathcal{F}(t)=u$
come in very narrow spikes. In other words, even if a time $t$ exists
for which the system returns close to its initial state, at a later
time $t+\Delta t$, with $\Delta t$ very small, the system is likely
to differ strongly form the initial state, and be nearly orthogonal
to it if $\overline{\mathcal{F}}$ is small. 

\textbf{5)} Eq.~(\ref{eq:TR}) has been derived assuming a finite
number $d$ of non-zero $p_{j}$. Nothing really dangerous happens
taking the limit $d\to\infty$, provided the spectrum stays discrete
and numerable and $\Delta E$ and $\overline{\mathcal{F}}$ stay finite
and non-zero ($\overline{\mathcal{F}}$ is finite because $\overline{\mathcal{F}}\le\langle\psi|\psi\rangle=1$).
In fact also in this case $\mathcal{F}(t)$ is an almost periodic
function. In a more general situation $H$ can also have a continuous
spectrum. The characteristic function $\chi(t)=\langle\psi|\psi(t)\rangle$
can then be written as
\begin{equation}
\chi(t)=\sum_{n}p_{n}e^{-iE_{n}t}+\int g(\epsilon)e^{-it\epsilon}d\epsilon,\label{eq:chi_cont}
\end{equation}
where the positive function $g(\epsilon)$ (this is the continuous
part of the projected density of states, PDOS, $\langle\psi|\delta(H-\epsilon)|\psi\rangle$)
has support on the continuous spectrum. If $g(\epsilon)d\epsilon$
is absolutely continuous, the Riemann-Lebesgue lemma guarantees that
$\lim_{t\to\infty}\int g(\epsilon)e^{-it\epsilon}d\epsilon=0$ . Hence,
in this case, the second term of Eq.~(\ref{eq:chi_cont}) contributes
at most a finite number of zeroes. A similar argument with small modifications,
works for the fidelity $\left|\chi(t)\right|^{2}$. Hence we deduce
that Eq.~(\ref{eq:TR}) works also in this more general setting with
the caveat that the $p_{n}$s refer only to the discrete spectrum
and one may not have $\sum_{n}p_{n}=1$. Note that Eq.~(\ref{eq:TR})
should be utilized with $u\in(0,M)$ with $M=\sum_{n}p_{n}<1$. Moreover
the Gaussian approximation employed may become worse if the missing
term $\int g(\epsilon)d\epsilon$ becomes a sensible fraction of $1.$

The above results inspire some connections for the debate around the
quantum speed limit if we consider Eq.~(\ref{eq:TR}) for $u\ll1$
\cite{mandelstam_uncertainty_1945,margolus_maximum_1998,giovannetti_quantum_2003,giovannetti_quantum_2003-1}.
In some situations (for example to estimate the computational power
of the ultimate laptop or the universe \cite{lloyd_ultimate_2000,lloyd_computational_2002})
one is interested in counting the number of total distinct (i.e.,
orthogonal) states that a ``system can pass through in a given period
of time'' \cite{margolus_maximum_1998}. Calling $T_{QSL}$ the minimum
time it takes for a system to go to an orthogonal state, it is known
that $T_{QSL}\ge(\pi/2)/(\langle H\rangle-E_{0})$. Accordingly, in
Ref.~\cite{margolus_maximum_1998}, it was estimated that the maximum
number of states, $N_{\mathrm{max}}$, a system can pass through in
a time interval $T$, is given by $N_{\mathrm{max}}=(2/\pi)(\langle H\rangle-E_{0})T$.
However, in this situation where the initial state is specified, there
is no need to compute $N_{\mathrm{max}}$, but rather $N_{T}(u)$
is the number we are after. Now, if $T$ is sufficiently large, the
number of times the state returns at fidelity $u$, is precisely given
by $N_{T}(u)\simeq T/T_{R}=T\Delta E/\mathcal{U}(u/\overline{\mathcal{F}})$.
Some considerations are in order. i) Exactly at $u=0$, as discussed
in \textbf{3)}, one has actually $N_{T}(u)=o(T)$ (i.e.~$\lim_{T\to\infty}N(u)/T=0$)
for RI energies when at least nine levels are populated; ii) The estimate
$N_{\mathrm{max}}$ of Ref.~\cite{margolus_maximum_1998} qualitatively
agrees with our $N(u)$ in the region where exponential and square
root terms are ineffective, i.e.~for $u/\overline{\mathcal{F}}\sim1$.
In fact in this region we have $N_{T}(u)\sim T\Delta E$. Note that
$u/\overline{\mathcal{F}}\sim1$ is likely the region where one would
want to use these results, since, as we have seen, typically $\overline{\mathcal{F}}$
is very small. However the relevant energy scale is not $(\langle H\rangle-E_{0})$,
(and not even the standard deviation $\sqrt{\langle H^{2}\rangle-\langle H\rangle^{2}}$),
but rather the standard deviation  $\Delta E$ computed with the ``squared''
distribution $\nu_{n}=p_{n}^{2}/\overline{\mathcal{F}}$.

\section{Integrable case}

The result Eq.~(\ref{eq:TR}) is valid assuming rational independence
of the many-body spectrum. This condition is massively violated for
quasi-free integrable models whose Hamiltonian is quadratic in creation
and annihilation Fermi/Bose operator. It is an interesting question
per se to compare typical timescale of integrable systems with non-integrable
ones %
\footnote{For instance it was shown that the equilibration time is generally
not too sensitive to integrability and is generally $O(J/\hbar)$
\cite{campos_venuti_unitary_2010,campos_venuti_equilibration_2013}.%
}. We consider hence a quasi-free system of fermions where the Hamiltonian
$H$ is bilinear in $c_{i},\, c_{i}^{\dagger}$ ($c_{j}$ Fermi annihilation
operators) and the initial state is Gaussian (i.e.~satisfies Wick's
theorem). For simplicity we consider a one-dimensional geometry and
assume that the fidelity can be written in the form 
\begin{equation}
\mathcal{F}(t)=\prod_{k}\left[1-\alpha_{k}\sin^{2}(t\epsilon_{k}/2)\right],\label{eq:LE_int}
\end{equation}
 where $\{\epsilon_{k}\}$ is the one-particle spectrum of $H$, $k$
is a quasi-momentum label, i.e.~$k=\pi(2n+1)/L$, with $n=0,1,\ldots,L$,
$L$ is the number of sites and $\alpha_{k}\in[0,1]$. This is the
case for example for quenches of the one dimensional XY model, but
can be valid more generally provided the Hamiltonian and the state
are translationally invariant. In particular generalization of Eq.~(\ref{eq:LE_int})
to $D$-dimensions is straightforward promoting $k$ to a $D$-dimensional
vector in the Brillouin zone. 

This is one of the rare cases where the calculation is harder for
integrable systems because the need of passing to the one particle
space introduces additional complications. For quasi-free systems
it is known that the logarithm of the fidelity (instead of the fidelity
itself) becomes Gaussian distributed \cite{campos_venuti_exact_2011}.
It is then natural to consider the variable $Z(t)=\log\mathcal{F}(t)$.
We will then estimate the number of zeroes of the equation $\log\mathcal{F}(t)=\ln u$.
Proceeding as previously we are now led to consider the joint probability
distribution $P_{Z,Z'}(z,z')=\overline{\delta(Z(t)-z)\delta(Z'(t)-z')}$.
For large $L$, $P_{Z,Z'}$ becomes Gaussian and using Eq.~(\ref{eq:Du})
one obtains
\begin{equation}
T_{R}=\frac{\pi\sigma_{Z}}{\sigma_{Z'}}\exp\left[(\ln u-\overline{\ln\mathcal{F}})^{2}/(2\sigma_{Z}^{2})\right],\label{eq:TR_integrable}
\end{equation}
with coefficients given by 
\begin{eqnarray}
\overline{\ln\mathcal{F}} & = & \sum_{k}\overline{z_{k}}\label{eq:corr_sum1}\\
\sigma_{Z}^{2} & = & \sum_{k}\left[\overline{(z_{k})^{2}}-(\overline{z_{k}})^{2}\right]\\
\sigma_{Z'}^{2} & = & \sum_{k}\overline{(z'_{k})^{2}},\label{eq:corr_sum3}
\end{eqnarray}
and $\overline{z_{k}},\,\overline{(z_{k})^{2}},\,\overline{(z'_{k})^{2}}$
are smooth, bounded, function of $\alpha_{k},\epsilon_{k}$. Some
remarks are in order.

\textbf{1)} The sum over $k$ in Eqns.~(\ref{eq:corr_sum1})-(\ref{eq:corr_sum3}),
runs over an extensive number of terms. Accordingly (since the summand
functions are bounded) we conclude that both $\overline{\ln\mathcal{F}}$
and $\sigma_{F}^{2}$ are extensive quantity. This implies that the
recurrence time for quasi-free fermions is only exponentially large
in the system volume. 

\textbf{2)} The behavior at the border $u\to0,1$ must be handled
with particular care. Similarly as in the non-integrable case, Eq.~(\ref{eq:TR_integrable})
predicts a finite density of exact recurrences times and a zero density
of orthogonalization times. The small, finite density of exact recurrences
predicted by Eq.~(\ref{eq:TR_integrable}) at $u=1$, is a consequence
of the Gaussian approximation. In fact a similar argument as for the
non-integrable case shows that, if all $\alpha_{k}$ are positive,
$\mathcal{F}(t)=1$ only at $t=0$. 

\textbf{3)} The exponential vs.~the double exponential dependence
of the recurrence time with the system's volume may suggest that $T_{R}$
may serve as a detector of integrability. However, any (no matter
how small) non-integrable perturbation of an integrable model will
result (with probability one) in a rationally independent many-body
spectrum and consequently a $T_{R}$ given by Eq.~(\ref{eq:TR}).
As a consequence $T_{R}$ will be discontinuous at the integrable
point. A similar behavior has been observed for the temporal variances
\cite{campos_venuti_gaussian_2013}.

\section{Fast recurrences\label{sec:Numerical-experiment}}

\begin{figure}
\noindent \begin{centering}
\includegraphics[width=4cm,height=2.8cm]{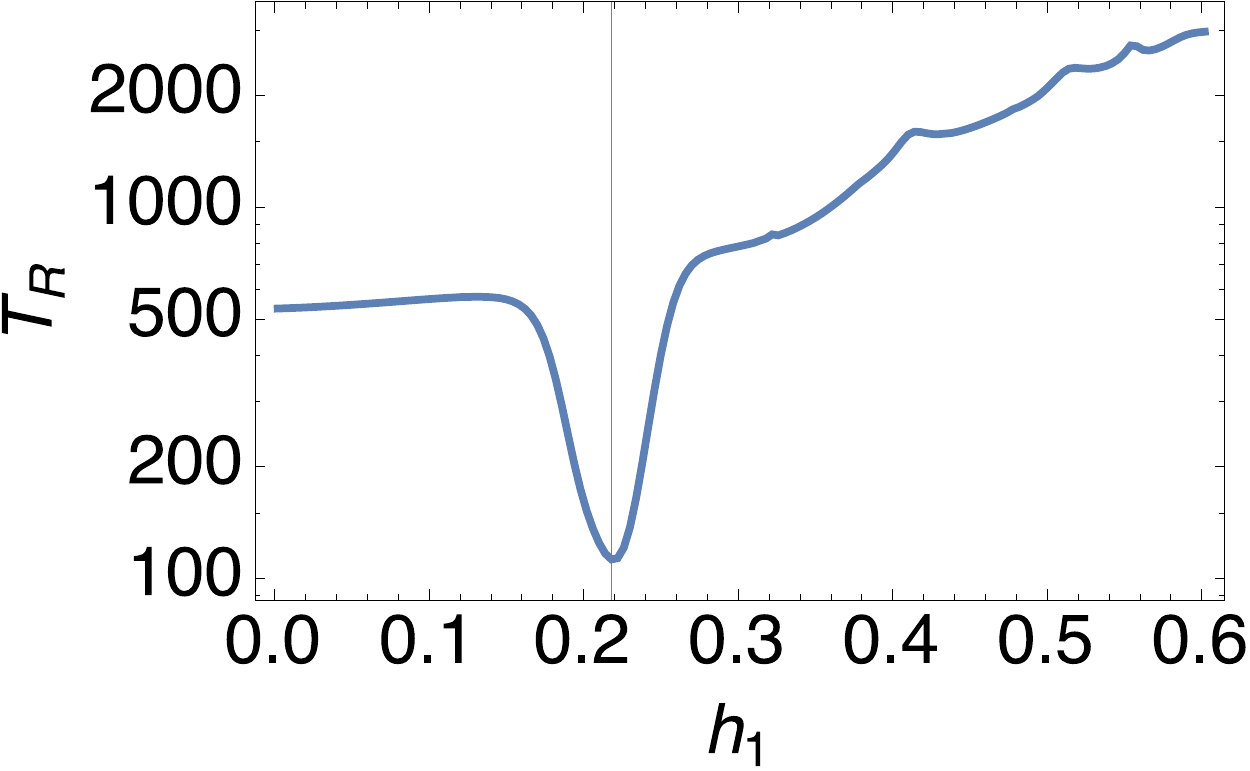}~~~\includegraphics[width=4cm,height=3cm]{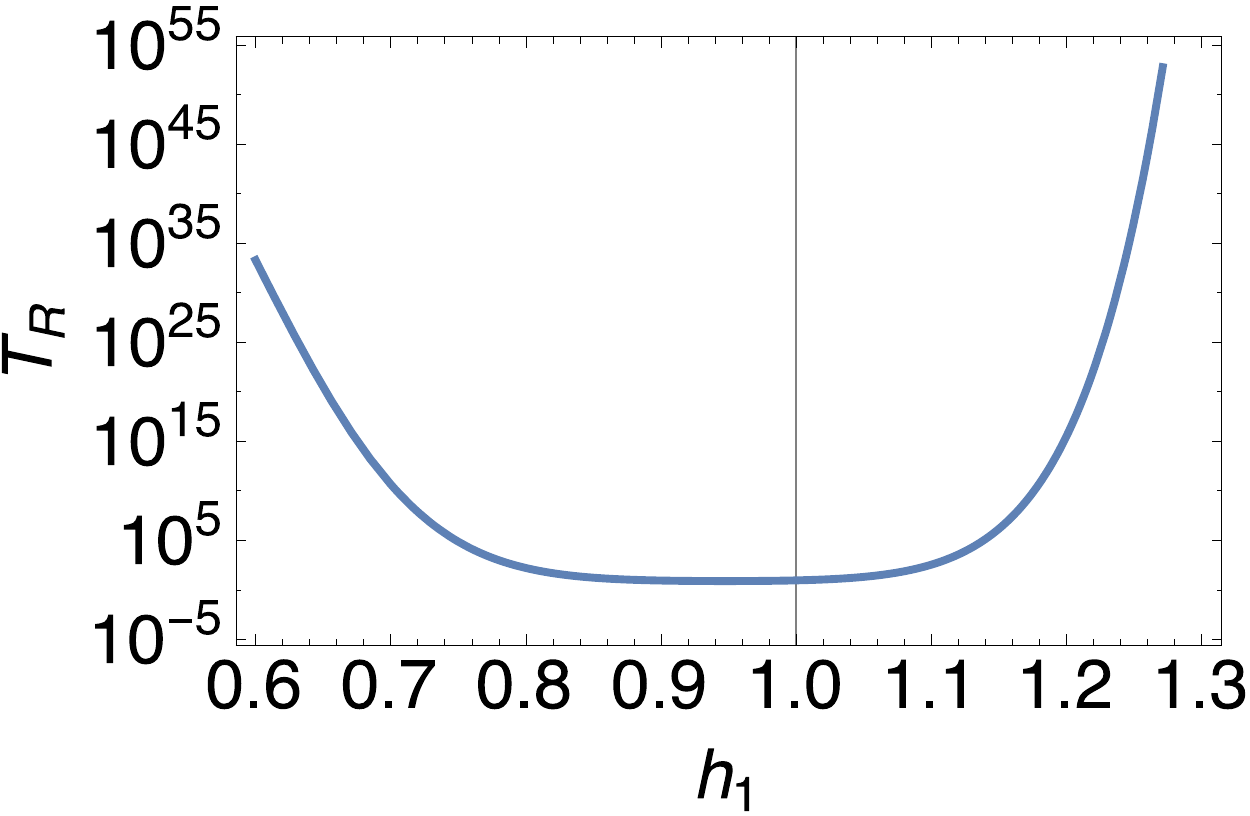}
\par\end{centering}

\protect\caption{Recurrence time in small quench experiments near criticality. The
system is initialized in the ground state of Hamiltonian (\ref{eq:TAM})
with parameter $\kappa_{1},\, h_{1}$ and then let evolve with the
same Hamiltonian with parameters $\kappa_{2},\, h_{2}$. Left panel:
non-integrable case with $\kappa_{1}=\kappa_{2}=0.4$. Other parameters
are $L=12,\,,h_{2}=h_{1}+\delta h,\delta h=0.03,u=0.98$. The recurrence
time is given by Eq.~(\ref{eq:TR}). The vertical line signals the
estimated critical point $h_{c}=0.218$. Right panel: integrable case
with $\kappa_{1}=\kappa_{2}=0$. Other parameters are $L=12,\,,h_{2}=h_{1}+\delta h,\delta h=0.03,u=0.98$.
The recurrence time is given by Eq.~(\ref{eq:TR_integrable}). The
vertical line signals the critical point $h_{c}=1$.\label{fig:small_quench}}
\end{figure}

The double exponential growth of the recurrence time with the system's
volume, poses serious questions on the possibility of observing recurrence
times in practical situations. On the other hand in some cases it
may be possible to prepare the initial state such that $p_{n}=0$
but for a few terms. Rabi oscillations may be seen as a limiting case
of this situation. Another possibility is provided by a small quench
experiment. In a quench experiment the system is initialized in the
ground state of the system's Hamiltonian with certain parameters $\lambda_{1}$.
The parameters are then suddenly changed and the system is evolved
with parameters $\lambda_{2}=\lambda_{1}+\delta\lambda$. Whereas
a general perturbation results in roughly as many excitations as the
Hilbert's space dimension, for sufficiently small quench amplitude
$\delta\lambda$, the number of excited quasi-particles is proportional
to the system's volume. A sufficient condition for small quench is
given by $\left|\delta\lambda\right|\ll L^{-D/2}$ where $L$ is the
linear system's size and the $D$ spatial dimensionality. As a consequence
the system evolves effectively in a small Hilbert space whose dimension
is roughly given by $1/\overline{\mathcal{F}}$ \cite{linden_quantum_2009}
and the recurrence time can be diminished by several order of magnitudes.
In fact one has $\overline{\mathcal{F}}\simeq1-\delta\lambda^{2}cV$
($c$ positive constant $O(1)$) and hence $T_{R}\simeq e^{u\delta\lambda^{2}cV}$.
However a further reduction is possible. If the system is close to
a critical point, the effective dimension $1/\overline{\mathcal{F}}$
is brought down to $2-3$ and $T_{R}$ becomes roughly $J^{-1}O(1)$.
A sufficient condition for small quench becomes in this case $\left|\delta\lambda\right|\ll L^{-1/\nu}$
where $\nu$ is the critical exponent of the correlation length $\xi$
\cite{campos_venuti_theory_2015-1}, whereas the quasi-critical regime
is defined by by $\xi(\lambda_{i})\gg L$. In this regime the recurrence
time is roughly size independent and may well become observable. 

To illustrate these effects we show numerical results for the one-dimensional
Ising model in transverse field with additional next nearest neighbor
interaction. The, so called, TAM Hamiltonian, is given by
\begin{equation}
H=-\sum_{i=1}^{L}\left(\sigma_{i}^{x}\sigma_{i+1}^{x}-\kappa\sigma_{i}^{x}\sigma_{i+2}^{x}+h\sigma_{i}^{z}\right),\label{eq:TAM}
\end{equation}
with periodic boundary conditions ($\sigma_{L+i}^{x}=\sigma_{i}^{x}$).
Hamiltonian Eq.~(\ref{eq:TAM}) is integrable at $\kappa=0$ and
non-integrable for all $\kappa\neq0$ (see e.g.~\cite{beccaria_density-matrix_2006,beccaria_evidence_2007}).
The parameters are initialized to $\kappa_{1},\, h_{1}$ and then
suddenly changed to $\kappa_{2},\, h_{2}$. We first illustrate the
(double exponential) reduction in recurrence time for a small quench
close to a non-integrable critical point. For small frustration $\kappa\le1/2$,
there is a transition from a ferromagnetic to a paramagnetic phase
increasing the external field $h$. Numerical simulations using Eq.~(\ref{eq:TR})
are shown in Fig.~\ref{fig:small_quench} left panel. It is evident
the sharp drop close to the quantum critical point. 

Similar considerations also apply to the integrable case for which
we set $\kappa=0$ in Eq.~(\ref{eq:TAM}). In this case the fidelity
is precisely given by Eq.~(\ref{eq:LE_int}) %
\footnote{Explicitly $\alpha_{k}=\sin^{2}(\vartheta_{k}^{(2)}-\vartheta_{k}^{(1)})$,
$\epsilon_{k}=2\sqrt{\sin^{2}(k)+(h_{2}+\cos(k))^{2}}$ and the Bogoliubov
angles given by $\tan\vartheta_{k}^{(i)}=-\sin(k)/(h_{i}+\cos(k))$. %
}. Again one observes a drop (albeit less sharp) close to the quantum
critical point Fig.~\ref{fig:small_quench} right panel.

\section{Conclusions}

We provided essentially exact formulas for the average recurrence
time in quantum systems both for general non-integrable and integrable
models. This is the --average-- time a system takes to get back to
the initial state up to an error $1-u$ in fidelity. We have shown
that, in the typical case, recurrence times are doubly exponential
in $V$, i.e.~$T_{R}\sim(\hbar/J)e^{ue^{\alpha V}}$, where $V$
is the dimensionless system's volume normalized by the volume of the
unit cell, $J$ a system's energy scale and $\alpha$ a positive constant.
Most of the time a valid approximation is given by $T_{R}\sim(\hbar/J)e^{ue^{S}}$
where $S$ is the Von Neumann's entropy of the equilibrium state.
For integrable systems instead the recurrence times are down to a
simple exponential in $V$. 

A possibility to drastically reduce such astronomical recurrence times
is to perform a small quench experiment. By this we mean preparing
the system in the ground state; slightly change some Hamiltonian parameters,
and let the system evolve undisturbed thereafter. In this situation
recurrences happen on a time-scale which is only exponential in the
volume. Furthermore if the quench is performed close to a quantum
critical point recurrence times become roughly size independent. This
drastic reduction opens up the possibility to experimentally observe
such fast recurrences in several nearly isolated quantum platforms.
In particular ion traps with order of $10$ atoms seems to be particularly
suited. 
\begin{acknowledgments}
The author would like to thank Leonardo Banchi and Larry Goldstein
for interesting discussions. This work was supported under ARO MURI
Grant No.~W911NF-11-1-0268. 
\end{acknowledgments}

\bibliographystyle{apsrev4-1}
\bibliography{poincare_time}

\appendix

\section{Densities of zeroes in the non-integrable case\label{sec:Densities-noint}}

The fidelity is the modulus square of the characteristic function
\begin{equation}
\chi(t)=\sum_{n}p_{n}e^{-itE_{n}}\,,\label{eq:chi}
\end{equation}
with weights $p_{n}=\langle\psi|\Pi_{n}|\psi\rangle$. The investigation
of sums of the form Eq.~(\ref{eq:chi}) was initiated long ago by
Lagrange \cite{lagrange_theorie_1781}. The study of the variation
of perihelion leads to a study of the variation of the argument of
such a sum where the number of terms is the number of planets. Here
instead we are interested in the behavior of the modulus. 

We now write $\chi=X+iY$ such that $\mathcal{F}=X^{2}+Y^{2}$ and
$\mathcal{F}'=2XX'+2YY'$ with
\begin{eqnarray}
X & = & \sum_{j}p_{j}\cos(E_{j}t)\label{eq:X}\\
Y & = & \sum_{j}p_{j}\sin(E_{j}t)\label{eq:Y}\\
X' & = & -\sum_{j}p_{j}E_{j}\sin(E_{j}t)\label{eq:X1}\\
Y' & = & \sum_{j}p_{j}E_{j}\cos(E_{j}t).\label{eq:Y1}
\end{eqnarray}
Note that, to avoid an overburdened notation, we use the same latter
to indicate both a function of time (e.g.~$X(t)$) and a random variable
($X$) obtained convoluting $t$ seen as a random variable with uniform
distribution in $[0,T]$ and taking the limit $T\to\infty$. We also
use the compact notation $\boldsymbol{X}=(X,Y,X',Y')$ and $\boldsymbol{x}=(x,y,x',y')$.
With the help of the joint distribution function $P_{\boldsymbol{X}}(\boldsymbol{x})$,
the density of zeros can be written as
\begin{multline}
D(u)=\int dxdydx'dy'\,\delta(u-x^{2}-y^{2})2\left|xx'+yy'\right|\times\\
P_{\boldsymbol{X}}(x,y,x',y').\label{eq:Du_multi}
\end{multline}
From equations (\ref{eq:X})--(\ref{eq:Y1}) and the RI assumption
it follows that $\boldsymbol{X}$ is a sum of independent random vectors.
Indeed the characteristic function of $\boldsymbol{X}$ ($\boldsymbol{k}=(\xi,\eta,\xi',\eta')$)
is
\begin{align}
\varphi_{\boldsymbol{X}}(\boldsymbol{k}) & =\overline{e^{i\boldsymbol{k}\cdot\boldsymbol{X}}}\\
 & =\prod_{j}\int\frac{d\vartheta_{j}}{2\pi}\exp[iA_{j}\cos(\vartheta_{j})+iB_{j}\sin(\vartheta_{j})]\\
 & =\prod_{j}J_{0}(\sqrt{A_{j}^{2}+B_{j}^{2}})=e^{-\sum_{j}(A_{j}^{2}+B_{j}^{2})/4+O(\boldsymbol{k}^{4})},\label{eq:Bessel_prod}
\end{align}
where $J_{0}$ is the Bessel function of the first kind and with $A_{j}=p_{j}(\xi+E_{j}\eta')$
and $B_{j}=p_{j}(\eta-E_{j}\xi')$. Using Eq.~(\ref{eq:Bessel_prod})
it may be possible to rigorously prove normality of (a properly rescaled
version of) $\boldsymbol{X}$ under some conditions using the central
limit theorem for triangular arrays. This would require to have a
well defined family of models parametrized, for example, by the size.
This is not always possible and so it will not be further pursued
here. Discarding the small, $O(\boldsymbol{k}^{4})$, term in Eq.~(\ref{eq:Bessel_prod})
we realize that $P_{\boldsymbol{X}}(\boldsymbol{x})$ has approximately
the following Gaussian form
\begin{eqnarray}
P_{\boldsymbol{X}}(\boldsymbol{x}) & = & \frac{e^{-\boldsymbol{x}^{T}\Sigma^{-1}\boldsymbol{x}/2}}{\pi^{2}\Delta},\label{eq:P}
\end{eqnarray}
with inverse correlation matrix given by
\begin{equation}
\Sigma^{-1}=\frac{2}{(DF-E^{2})}\left(\begin{array}{cccc}
F & 0 & 0 & -E\\
0 & F & E & 0\\
0 & E & D & 0\\
-E & 0 & 0 & D
\end{array}\right),
\end{equation}
having defined
\begin{eqnarray}
D & = & \sum_{j}p_{j}^{2}=\overline{\mathcal{F}}\\
E & = & \sum_{j}p_{j}^{2}E_{j}\\
F & = & \sum_{j}p_{j}^{2}E_{j}^{2}\,.
\end{eqnarray}
The positivity of the matrix $\Sigma$ reduces to $\Delta:=DF-E^{2}\ge0$
which can be easily proved using Cauchy-Schwarz inequality with the
vectors $\boldsymbol{v}=(p_{1},p_{2},\ldots)$ and $\boldsymbol{w}=(E_{1}p_{1},\omega_{2}p_{2},\ldots)$
(under assumption of convergence) $\left|\boldsymbol{v}\cdot\boldsymbol{w}\right|^{2}\le\left\Vert \boldsymbol{v}\right\Vert ^{2}\left\Vert \boldsymbol{w}\right\Vert ^{2}$.

We now pass to polar coordinates $\left(x,y\right)=\rho_{1}\left(\cos\vartheta_{1},\sin\vartheta_{1}\right)$
and $\left(x',y'\right)=\rho_{2}\left(\cos\vartheta_{2},\sin\vartheta_{2}\right)$.
The density of zeroes becomes 
\begin{align}
D(u) & =\int dxdydx'dy'\,\delta(u-x^{2}-y^{2})2\left|xx'+yy'\right|\nonumber \\
 & \times P_{\boldsymbol{X}}(x,y,x',y')\label{eq:I_main}\\
 & =\int_{0}^{\infty}\!\! d\rho_{1}\int_{0}^{\infty}\!\! d\rho_{2}\,(\rho_{1}\rho_{2})^{2}\delta(u-\rho_{1}^{2})P_{\boldsymbol{X}}(\boldsymbol{\rho},\boldsymbol{\vartheta})\\
 & \times\int_{0}^{2\pi}\!\! d\vartheta_{1}\int_{0}^{2\pi}\!\! d\vartheta_{2}\,2\left|\cos(\vartheta_{1}-\vartheta_{2}))\right|.
\end{align}

Using the Gaussian approximation Eq.~(\ref{eq:P}), the integration
over $\rho_{1}$ is trivial and the result is
\begin{multline}
D(u)=\sqrt{u}\int_{0}^{\infty}\!\! d\rho_{2}\rho_{2}^{2}\int_{0}^{2\pi}\!\! d\vartheta_{1}\int_{0}^{2\pi}\!\! d\vartheta_{2}\\
\times\left|\cos(\vartheta_{1}-\vartheta_{2})\right|\tilde{P}(\rho_{2},\vartheta_{1},\vartheta_{2})\label{eq:D-exact-1}
\end{multline}
 where 
\[
\tilde{P}(\rho_{2},\vartheta_{1},\vartheta_{2})=e^{-\alpha\rho_{2}^{2}-\beta\sin(\vartheta_{1}-\vartheta_{2})\rho_{2}-\gamma}\frac{1}{\pi^{2}\Delta}
\]
and the coefficients are fixed by
\begin{multline*}
\left.\boldsymbol{x}\Sigma^{-1}\boldsymbol{x}/2\right|_{\rho_{1}=\sqrt{u}}=\\
\frac{D\rho_{2}^{2}+2E\sqrt{u}\rho_{2}\sin(\vartheta_{1}-\vartheta_{2})+Fu}{DF-E^{2}},
\end{multline*}
and are given by
\begin{align}
\alpha & =\frac{D}{\Delta}\\
\beta & =\frac{2E\sqrt{u}}{\Delta}\\
\gamma & =\frac{Fu}{\Delta}.
\end{align}

Now the angular integration can be performed
\begin{multline}
\int_{0}^{2\pi}\!\! d\vartheta_{1}\int_{0}^{2\pi}\!\! d\vartheta_{2}\,\left|\cos(\vartheta_{1}-\vartheta_{2})\right|e^{-\beta\rho_{2}\sin(\vartheta_{1}-\vartheta_{2})}\\
=8\pi\frac{\sinh(\beta\rho_{2})}{\beta\rho_{2}}.
\end{multline}
The remaining integration over $\rho_{2}$ is Gaussian and the final
result is extremely simple
\begin{equation}
D(u)=\frac{2}{\sqrt{\pi}}\frac{\sqrt{u\Delta}}{\overline{\mathcal{F}}^{3/2}}e^{-u/\overline{\mathcal{F}}}.\label{eq:dens}
\end{equation}
Using the ensemble $\nu_{n}=p_{n}^{2}/\sum_{n}p_{n}^{2}$, the variance
of the energy $\{E_{n}\}$ is $\Delta E^{2}=[\sum_{j}p_{j}^{2}E_{j}^{2}/\overline{\mathcal{F}}-(\sum_{j}p_{j}^{2}E_{j}/\overline{\mathcal{F}})^{2}]=\Delta/\overline{\mathcal{F}}^{2}$
and we obtain
\begin{equation}
D(u)=\frac{2}{\sqrt{\pi}}\Delta E\sqrt{\frac{u}{\overline{\mathcal{F}}}}e^{-u/\overline{\mathcal{F}}},\label{eq:density}
\end{equation}
which, upon remembering that $T_{R}=1/D(u)$, coincides with Eq.~(\ref{eq:TR}).
The exponential term in Eq.~(\ref{eq:TR}) could have been obtained
with the following (wrong) argument sometimes used in physics. According
to this argument, $T_{R}(u)$, i.e., the average of the solutions
of $\mathcal{F}(t)=u$, is approximately estimated by the inverse
probability (density) of the event $\mathcal{F}(t)=u$ (time some
unknown time-scale $J^{-1}$). The probability density of the fidelity
has been computed in \cite{campos_venuti_unitary_2010,campos_venuti_universal_2014}
and is given by $P_{\mathcal{F}}(u)=\vartheta(u)\exp(-u/\overline{\mathcal{F}})/\overline{\mathcal{F}}$
($\vartheta(u)$ Heaviside function). This argument then predicts,
$T_{R}(u)\sim J^{-1}\overline{\mathcal{F}}\exp(u/\overline{\mathcal{F}})$
which fails if $u$ is not sufficiently away from zero.

\section{Proof of zero density for $u=0$\label{sec:zero_density}}

Throughout this section we assume that only a finite number $d'$
of $p_{n}$'s are non-zero. The result (\ref{eq:dens}) predicts a
zero density of orthogonalization times, i.e.~times for which $\mathcal{F}(t)=0$.
It is legitimate to ask weather this holds exactly or is an artifact
of the Gaussian approximation used to derive Eq.~(\ref{eq:dens}).
We then go back to equation (\ref{eq:I_main}) in its exact form that
we re-write here for clarity
\begin{eqnarray}
D(u) & = & \sqrt{u}\int_{0}^{\infty}d\rho_{2}\rho_{2}^{2}\int_{0}^{2\pi}d\vartheta_{1}\int_{0}^{2\pi}d\vartheta_{2}\,\left|\cos(\vartheta_{1}-\vartheta_{2})\right|\nonumber \\
 &  & \times P_{\boldsymbol{X}}(\sqrt{u}\cos\vartheta_{1},\sqrt{u}\sin\vartheta_{1}\rho_{2}\cos\vartheta_{2},\rho_{2}\sin\vartheta_{2})\nonumber \\
 & =: & \sqrt{u}R(u)\label{eq:uRu}
\end{eqnarray}
We will show that $R(u)$ is bounded for all $u$, in particular for
$u=0$. The result then simply follows from Eq.~(\ref{eq:uRu}) taking
the limit $u\to0$. Let us analyze the logic. First we show that,
by construction, $P_{\boldsymbol{X}}(\boldsymbol{x})$ is compactly
supported. Then we prove that $P_{\boldsymbol{X}}(\boldsymbol{x})$
is also bounded thus implying that $R(u)$ is bounded. 
\begin{description}
\item [{Lemma}] The function $P_{\boldsymbol{X}}(x,y,x',y')$ is compactly
supported provided $\langle\left|H\right|\rangle$ is finite, more
precisely, $P_{\boldsymbol{X}}(x,y,x',y')=0$ outside the (hyper)
rectangle $\left|x\right|,\left|y\right|\le1$ and $\left|x'\right|,\left|y'\right|\le C=\sum_{j}p_{j}\left|E_{j}\right|=\langle\left|H\right|\rangle$.
The latter quantity is certainly finite for the finite dimensional
case, physically it can be postulated to grow as the system's volume. 
\end{description}
It is obvious that $P_{\boldsymbol{X}}(x,y,x',y')=0$ for $\left|x\right|,\left|y\right|>1$.
On the other hand $\left|X'\right|=\left|-\sum_{j}p_{j}E_{j}\cos(E_{j}t)\right|\le\sum_{j}p_{j}\left|E_{j}\right|=C$,
QED. Now $P_{\boldsymbol{X}}(\boldsymbol{x)}$ is bounded provided
the characteristic function (its Fourier transform) $\varphi_{\boldsymbol{X}}(\boldsymbol{k})\in L^{1}(\mathbb{R}^{4}).$
The reminder of this section is devoted to asses the conditions under
which $\varphi_{\boldsymbol{X}}(\boldsymbol{k})$ is summable. 

Remind the definition of $\varphi_{\boldsymbol{X}}(\boldsymbol{k})$

\begin{equation}
\varphi_{\boldsymbol{X}}(\boldsymbol{k})=\prod_{j=1}^{d'}J_{0}(\sqrt{\boldsymbol{k}^{T}M_{j}\boldsymbol{k}})
\end{equation}
where $d'$ is the number of non-zero $p_{j}$ and 
\begin{equation}
M_{j}=p_{j}^{2}\left(\begin{array}{cccc}
1 & 0 & 0 & E_{j}\\
0 & 1 & -E_{j} & 0\\
0 & -E_{j} & E_{j}^{2} & 0\\
E_{j} & 0 & 0 & E_{j}^{2}
\end{array}\right).
\end{equation}
The matrix $M_{j}$ is hermitian, positive semi-definite, has eigenvalues
$0([\times2]$ and $p_{j}^{2}(1+E_{j}^{2})[\times2]$ and a null space
$V_{j}=\mathrm{span}\left\{ \left(E_{j},0,0,-1\right),\left(0,E_{j},1,0\right)\right\} $.
The Bessel function is smooth and everywhere bounded and hence so
is $\varphi_{\boldsymbol{X}}(\boldsymbol{k})$. To see wether it is
summable we must look at its behavior as $\boldsymbol{k}\to\infty$.
Remember that, for $x\to\infty$, neglecting an oscillating factor,
we have $\left|J_{0}(x)\right|\sim\sqrt{2/\pi}/\sqrt{x}$. Let us
fix an $\varepsilon>0$ and define the cylinders (in this picture
the planes become lines) $T_{j}=\{\boldsymbol{k}|\boldsymbol{k}M_{j}\boldsymbol{k}\le\epsilon^{2}\}$.
We call ``Tubes'' the collection (union) of all the $T_{j}$'s.
We want to see if and when the following integral is convergent
\begin{multline}
\int_{\mathbb{R}^{4}}\left|\varphi_{\boldsymbol{X}}(\boldsymbol{k})\right|d\boldsymbol{k}=\\
\int_{\mathbb{R}^{4}\backslash\mathrm{Tubes}}\left|\varphi_{\boldsymbol{X}}(\boldsymbol{k})\right|d\boldsymbol{k}+\int_{\mathrm{Tubes}}\left|\varphi_{\boldsymbol{X}}(\boldsymbol{k})\right|d\boldsymbol{k}.
\end{multline}
Let us call $I_{1}$ ($I_{2}$) the first (respectively, the second)
integral above. For clarity we also go to spherical coordinates $\boldsymbol{k}=(\rho,\Omega)$
and define $\sqrt{\boldsymbol{k}^{T}M_{j}\boldsymbol{k}}=\rho f_{j}(\Omega)$.
Note that $f_{j}(\Omega)=0$ on $V_{j}$ but $f_{j}(\Omega)>0$ on
$\mathbb{R}^{4}\backslash\mathrm{Tubes}$. Let us first consider $I_{1}$,
\begin{equation}
I_{1}=\int_{\mathbb{R}^{4}\backslash\mathrm{Tubes}}\prod_{j=1}^{d'}\left|J_{0}\left(\rho f_{j}(\Omega)\right)\right|d\boldsymbol{k}.
\end{equation}
To estimate convergence of the above we look at the integrand when
$\rho\to\infty$. We use then the asymptotic expansion of the Bessel
function and obtain (apart from unimportant constants and, bounded,
oscillating factors)
\begin{equation}
\prod_{j=1}^{d'}\left|J_{0}\left(\rho f_{j}(\Omega)\right)\right|\stackrel{\rho\to\infty}{\sim}\frac{1}{\rho^{d'/2}}\prod_{j=1}^{d'}\frac{1}{\sqrt{f_{j}(\Omega)}}.\label{eq:series_notubes}
\end{equation}
Note that the expansion in Eq.~(\ref{eq:series_notubes}) is well
defined because $f_{j}(\Omega)$ is never zero in the domain of integration.
Since we are in four dimensions we deduce that $I_{1}$ is convergent
if $d'/2>4$, i.e., $d'>8$. Let us then look at $I_{2}$. By sub-additivity
of the Lebesgue integral
\begin{eqnarray}
I_{2} & = & \int_{T_{1}\cup T_{2}\cdots}\left|\varphi_{\boldsymbol{X}}(\boldsymbol{k})\right|d\boldsymbol{k}\le\sum_{j=1}^{d'}\int_{T_{j}}\left|\varphi_{\boldsymbol{X}}(\boldsymbol{k})\right|d\boldsymbol{k}\\
 & \equiv & \sum_{j=1}^{d'}I_{2,j}
\end{eqnarray}
Let us consider $I_{2,1}$, the arguments are the same for all $j$.
On $T_{1}$, since $\boldsymbol{k}^{T}M_{1}\boldsymbol{k}\le\epsilon^{2}$
, $J_{0}\left(\sqrt{\boldsymbol{k}^{T}M_{1}\boldsymbol{k}}\right)\simeq1$
for small $\epsilon$, hence we use the simple bound $\left|J_{0}(x)\right|\le1$
and obtain
\begin{equation}
I_{2,1}\le\int_{T_{1}}\prod_{j=2}^{d'}J_{0}\left(\sqrt{\boldsymbol{k}^{T}M_{j}\boldsymbol{k}}\right).
\end{equation}
We also consider the change of variable $\boldsymbol{k}=U_{1}^{T}\boldsymbol{q}$
where $U_{1}$ is the matrix which diagonalizes $M_{1}$, i.e.~$U_{1}M_{1}U_{1}^{T}=D_{1}$.
Explicitly
\begin{equation}
U_{1}=\frac{1}{\sqrt{1+E_{1}^{2}}}\left(\begin{array}{cccc}
E_{1} & 0 & 0 & -1\\
0 & E_{1} & 1 & 0\\
1 & 0 & 0 & E_{1}\\
0 & 1 & -E_{1} & 0
\end{array}\right).
\end{equation}
 In these variables $T_{1}=\{\boldsymbol{q}|\, q_{3}^{2}+q_{4}^{2}\le\epsilon^{2}/[p_{1}^{2}(1+E_{1}^{2})]\}$.
We define $\tilde{M}_{j}=U_{1}M_{j}U_{1}^{T}$, so that 
\begin{equation}
I_{2,1}\le\int_{T_{1}}\prod_{j=2}^{d'}J_{0}\left(\sqrt{\boldsymbol{q}^{T}\tilde{M}_{j}\boldsymbol{q}}\right)d\boldsymbol{q}.
\end{equation}
For $\epsilon$ non-zero, there are points $\boldsymbol{q}$ in $T_{1}$
such that $\boldsymbol{q}^{T}\tilde{M}_{j}\boldsymbol{q}=0$, however
these regions are bounded. To see this note that the planes $V_{j}$s
intersects only at the origin. The thicked version of the $V_{j}$s,
$T_{j}$s, intersect only in a bounded region around the origin, because
the $\tilde{M}_{j}$ are positive semidefinite. In other words there
is no way to make $q_{1},q_{2}$ run to infinity while having $\boldsymbol{q}^{T}\tilde{M}_{j}\boldsymbol{q}=0$
such as to render the corresponding Bessel function ineffective. If
this argument is not convincing let us look directly at $\boldsymbol{q}^{T}\tilde{M}_{j}\boldsymbol{q}$.
With a further change of variables $\boldsymbol{q}=(\rho\cos(\vartheta),\rho\sin(\vartheta),R\cos(\phi),R\sin(\phi))$,
with Jacobian $R\rho$, we obtain
\begin{multline}
\boldsymbol{q}^{T}\tilde{M}_{j}\boldsymbol{q}=\frac{p_{j}^{2}}{1+E_{1}^{2}}\Big(R^{2}(1+E_{1}E_{J})^{2}+\rho^{2}(E_{1}-E_{j})^{2}\\
+2\left(E_{1}-E_{j}+E_{1}^{2}E_{j}-E_{1}E_{j}^{2}\right)R\rho\cos(\vartheta-\phi)\Big).
\end{multline}
For $\rho$ sufficiently large the above is never zero. In fact, for
large $\rho$, 
\begin{equation}
\boldsymbol{q}^{T}\tilde{M}_{j}\boldsymbol{q}=p_{j}^{2}\frac{(E_{1}-E_{j})^{2}}{1+E_{1}^{2}}\rho^{2}+O(\rho),
\end{equation}
such that, neglecting an oscillating factor, at leading order 
\[
\left|J_{0}\left(\sqrt{\boldsymbol{q}^{T}\tilde{M}_{j}\boldsymbol{q}}\right)\right|\stackrel{\rho\to\infty}{\longrightarrow}\sqrt{\frac{2\sqrt{1+E_{1}^{2}}}{\pi p_{j}\left|E_{1}-E_{j}\right|}}\frac{1}{\rho^{1/2}},
\]
whereas next to leading terms decay ever faster. All in all, neglecting
un-important constants, we are led to study the convergence of the
following integral (say for a certain $c>0$)
\[
\int_{0}^{2\pi}\!\! d\vartheta\int_{0}^{2\pi}\!\! d\phi\int_{0}^{\epsilon/\sqrt{p_{1}(1+E_{1}^{2})}}\!\! dR\int_{c}^{\infty}d\rho\, R\rho\,\frac{1}{\rho^{(d'-1)/2}}.
\]
The above is convergent for $(d'-1)/2>2$, i.e.~$d'>5$. The same
arguments can be repeated for all the $I_{2,j}$ implying that $I_{2}$
is convergent for $d'>5$. Overall is $\varphi_{\boldsymbol{X}}(\boldsymbol{k})$
is in $L^{1}(\mathbb{R}^{4})$ provided $d'>8$. Under this condition
we then have $\left|P_{\boldsymbol{X}}(\boldsymbol{x})\right|\le\int d\boldsymbol{k}\left|\varphi_{\boldsymbol{X}}(\boldsymbol{k})\right|<\infty$.
The result then follows noting that $\left|R(u)\right|\le C^{2}\sup_{\boldsymbol{x}}\left|P_{\boldsymbol{X}}(\boldsymbol{x})\right|$.

\section{Solutions of $\mathcal{F}(t)=1$\label{sec:LEeq1} }

We assume here that the discrete spectrum is finite and $\sum_{j=1}^{d}p_{j}=1$.
$\mathcal{F}(t)=1$ if and only if $\chi(t)=e^{-i\phi}$, in other
words, with $E_{n}'=E_{n}-\phi$
\begin{equation}
\sum_{j=1}^{d}p_{j}e^{-itE'_{n}}=1.
\end{equation}
Given the constraint $\sum_{j=1}^{d}p_{j}=1$, the above equation
can be satisfied if and only if all the phases are integer multiple
of $2\pi$, i.e.~there is a time $t$ for which $E'_{n}t=2\pi m_{n}$
with $m_{n}$ integers (depending on $t$). Assume $t\neq0$, in this
case $E'_{n}/m_{n}=2\pi/t=$constant. But this violates the assumption
of rational independence. In fact, if $d$ is even, simply take $d/2$
times rational coefficients $c_{n}=1/m_{n}$ and the other half, $c_{n}=-1/m_{n}$.
Then $\sum_{n=1}^{d}E'_{n}c_{n}=(d/2-d/2)2\pi/t=0$. For $d=2p+1$
odd take $p$ times $c_{n}=1/m_{n}$ and $p-1$ times $c_{n}=-1/m_{n}$.
For the remaining two cases pick $c_{n}=-1/(2m_{n})$. Then $\sum_{n=1}^{d}E'_{n}c_{n}=(p-(p-1)-1/2-1/2)2\pi/t=0$.
But the set $\{E_{n}'\}$ is rationally independent if the set $\{E_{n}\}$
is which is a contradiction. Hence $t=0$ is the only solution of
$\mathcal{F}(t)=1$.

\section{Density of zeroes for quasi-free fermions\label{sec:Density-integrable}}

Here we consider a system of quasi-free fermions whose fidelity is
given by Eq.~(\ref{eq:LE_int}). The first step is to find the joint
distribution $P_{Z,Z'}(z,z')=\overline{\delta(Z(t)-z)\delta(Z'(t)-z')}$.
Clearly
\begin{eqnarray}
Z(t) & = & \sum_{k}\ln\left[1-\alpha_{k}\sin^{2}(t\epsilon_{k}/2)\right]\\
Z'(t) & = & -\sum_{k}\frac{\alpha_{k}\epsilon_{k}\sin(t\epsilon_{k}/2)\cos(t\epsilon_{k}/2)}{1-\alpha_{k}\sin^{2}(t\epsilon_{k}/2)},
\end{eqnarray}
which we write as $Z=\sum_{k}z_{k}$, ($Z'=\sum_{k}z'_{k}$) for ease
of notation. We now assume rational independence of the one-particle
energies $\epsilon_{k}$. With this assumption all variables are independent
and uncorrelated except that $z_{k}$ ($z'_{k}$) correlates with
$z_{k}$ ($z'_{k}$). In fact it turns out that $\langle z'_{k}\rangle=0$
and $\langle z{}_{k}z'_{k}\rangle=0$ because $z{}_{k}$ is even while
$z'_{k}$ is an odd function of $t$. In general one expects a central
limit theorem and a Gaussian $P_{Z,Z'}(x_{1},x_{2})$. The only non-vanishing
correlations are 
\begin{eqnarray}
\overline{\ln\mathcal{F}} & = & \sum_{k}\overline{z_{k}}\\
\sigma_{Z}^{2} & = & \sum_{k}\left[\overline{(z_{k})^{2}}-(\overline{z_{k}})^{2}\right]\\
\sigma_{Z'}^{2} & = & \sum_{k}\overline{(z'_{k})^{2}}.
\end{eqnarray}
Time averages are translated into phase-space averages using the assumption
of rational independence of the one particle energies. Evaluating
the integrals one finds
\begin{eqnarray}
\overline{z_{k}} & = & 2\ln\frac{1+\sqrt{1-\alpha_{k}}}{2}\\
\overline{(z'_{k})^{2}} & = & \frac{\left(2-2\sqrt{1-\alpha_{k}}-\alpha_{k}\right)}{2\sqrt{1-\alpha_{k}}}\epsilon_{k}^{2}.
\end{eqnarray}

The expression for $\overline{(z_{k})^{2}}$ is slightly cumbersome.
The second moment can be written as
\begin{eqnarray}
\overline{(z_{k})^{2}} & = & \int_{0}^{2\pi}\left[\ln\left(1-\alpha_{k}\sin^{2}\left(\vartheta/2\right)\right)\right]^{2}\frac{d\vartheta}{2\pi}\\
 & = & \frac{2}{\pi}\int_{0}^{1}\frac{\left[\ln\left(1-\alpha_{k}y^{2}\right)\right]^{2}}{\sqrt{1-y^{2}}}dy,
\end{eqnarray}
where we changed variable in the second line . The last integral can
be expressed in terms of logarithms and dilogarithms. The final expression
for the variance $\sigma_{z_{k}}^{2}:=\overline{(z_{k})^{2}}-(\overline{z_{k}})^{2}$,
is
\begin{align}
\sigma_{z_{k}}^{2} & =-4\log^{2}\left(\sqrt{1-\alpha_{k}}+1\right)-4\log(2)\log\left(\alpha_{k}\right)\nonumber \\
 & +4\log\left(4-4\sqrt{1-\alpha_{k}}\right)\log\left(\sqrt{1-\alpha_{k}}+1\right)\nonumber \\
 & +4i\pi\log\left(\frac{2}{\sqrt{1-\alpha_{k}}+1}\right)+4\text{Li}_{2}\left(\frac{2\left(\sqrt{1-\alpha_{k}}+1\right)}{\alpha}\right)\nonumber \\
 & -4\text{Li}_{2}\left(\frac{-\alpha_{k}+2\sqrt{1-\alpha_{k}}+2}{\alpha_{k}}\right).
\end{align}

All in all, the joint distribution of $Z,\, Z'$ is given by
\[
P_{Z,Z'}(z,z')=\frac{1}{2\pi\sigma_{Z}\sigma_{Z'}}\exp\left[-\frac{(z-\overline{\ln\mathcal{F}})^{2}}{2\sigma_{Z}^{2}}-\frac{(z')^{2}}{2\sigma_{Z'}^{2}}\right].
\]
The evaluation of the densities of the zeroes of the equation $\ln\mathcal{F}(t)=\ln u$
proceeds similarly as in Sec.~. The evaluation of the integrals turns
out to be simpler because the variables $z_{k}$ and $z_{k}'$ are
factorized . The final result is
\begin{equation}
D(u)=\frac{\sigma_{Z'}}{\pi\sigma_{Z}}\exp\left[-(\ln u-\overline{\ln\mathcal{F}})^{2}/(2\sigma_{Z}^{2})\right],\label{eq:density_integrable}
\end{equation}
which gives Eq.~(\ref{eq:TR_integrable}).
\end{document}